\pdfoutput=1

\documentclass[11pt]{article}
\usepackage{jheppub}

\usepackage{amsmath}
\usepackage{amssymb}
\usepackage{graphicx,color,slashed,braket}
\usepackage{bbm}
\usepackage{ifpdf}
\usepackage{comment}

\usepackage{booktabs}
\usepackage{multirow}
\usepackage{makecell}
\usepackage{float}
\usepackage{verbatim}
\usepackage{caption}
\usepackage{cancel}
\usepackage{enumerate}
\usepackage{tcolorbox}
\usepackage[font=small,labelfont=bf]{caption}

  \usepackage{color}
  \definecolor{dark-gray}{gray}{0.20}
  \definecolor{gray}{gray}{0.30}
  \definecolor{light-gray}{gray}{0.80}
  \definecolor{dark-red}{rgb}{0.7,0,0}
  \definecolor{dark-green}{rgb}{0.1,0.4,0}
  \definecolor{dark-blue}{rgb}{0.3,0.3,0.7}
  \definecolor{light-blue}{rgb}{0.8,0.8,1}

\usepackage{hyperref}
\usepackage{setspace}
\hypersetup{
	colorlinks=true,
	linkcolor=dark-blue,
	citecolor=dark-red,
	urlcolor=dark-green,
	linktoc=page
}
\newcommand{\be}{\begin{equation}}
\newcommand{\ee}{\end{equation}}

\def\be{\begin{equation}}
\def\ee{\end{equation}}
\def\bea{\begin{eqnarray}}
\def\eea{\end{eqnarray}}

\renewcommand{\]}{\right]}
\renewcommand{\(}{\left(}
\renewcommand{\)}{\right)}

\newcommand{\vol}{\text{vol}}

\newcommand{\e}{\textrm{e}}

\newcommand{\dd}{\mathrm{d}}

\allowdisplaybreaks

\title{Scale-separated AdS$_3$ vacua from $G_2$-orientifolds using bispinors}

\author{Vincent Van Hemelryck}

\affiliation{Instituut voor Theoretische Fysica, KU Leuven,
Celestijnenlaan 200D, B-3001 Leuven, Belgium}

\emailAdd{vincent.vanhemelryck@kuleuven.be}

\notoc

\abstract{In this paper, minimal supersymmetric AdS$_3$ solutions on $G_2$-orientifolds of type IIA and IIB supergravity are studied. These are obtained by techniques of the bispinor formalism. A straightforward argument is derived against AdS$_3$ solutions in massless type IIA string theory on such $G_2$-backgrounds. Furthermore, it is shown that some of the AdS$_3$ solutions in massive type IIA do admit a separation of scales, whereas the same cannot be said for the type IIB vacua.
Finally, a formula for the 3d superpotential in the bispinor language is given that is valid for both type IIA and type IIB configurations. It is shown that the supersymmetry conditions of this superpotential lead back to the bispinor equations.}

\begin{document}

\maketitle
\newpage
\tableofcontents

\section{Introduction}
The swampland program sheds new light on the field of string compactifications. Its aim of delineating which low energy effective field theories (EFTs) can be completed into a consistent UV theory of quantum gravity and which not, has led to many different insights. However, the criteria for EFTs to belong to the Swampland are formulated as conjectures. Some of these conjectures have been extensively studied in the context of ten-dimensional string theory, e.g. the No Global Symmetries conjecture \cite{Banks:1988yz,Banks:2010zn}, the Weak Gravity Conjecture \cite{ArkaniHamed:2006dz}, etc. for which no counterexamples have been found and arguments close to a proof have been formulated \cite{Harlow:2018jwu, Harlow:2018tng,Montero:2018fns}. Unfortunately other conjectures, with more phenomenological impact, do not share the same fate. An example is the strong AdS distance conjecture (ADC) \cite{Lust:2019zwm}, which forbids supersymmetric AdS vacua that exhibit a feature called scale separation. Similar statements were already raised before in \cite{Gautason:2015tig, Gautason:2018gln} and more recently in \cite{Collins:2022nux}.
A string compactification is said to be scale-separated whenever its Kaluza-Klein (KK) energy scale is much larger than the cosmological constant scale. For phenomenological purposes, this feature is vital, as only scale-separated vacua can be considered genuinely lower-dimensional.
It has been shown for $d=4$ that scale separation is in tension with the magnetic weak gravity conjecture for $\mathcal{N}>1$ supersymmetric AdS$_d$ vacua purely from the lower-dimensional gauged supergravity point of view \cite{Cribiori:2022trc}. Similar arguments could rule out any supersymmetric $d>4$ theory, but has to be investigated further.
For $d \leq 4$ the situation is a bit more involved. Indeed, there are a couple of counterexamples to the ADC. In type IIB string theory one has e.g. the four-dimensional supersymmetric KKLT \cite{Kachru:2003aw} vacuum where scale separation is achieved by a small flux superpotential, where one has to rely on non-perturbative physics to stabilise the moduli. Finding vacua with such a small flux superpotential has been the topic of many recent works \cite{Demirtas:2019sip,Demirtas:2020ffz,Alvarez-Garcia:2020pxd,Demirtas:2021nlu,Demirtas:2021ote,Gendler:2022qof,Blumenhagen:2022dbo}, but has been criticised in \cite{Lust:2022lfc, Blumenhagen:2022dbo}. Another class of models with apparent scale separation is the (non-supersymmetric) Large Volume Scenario (LVS) \cite{Balasubramanian:2005zx,Conlon:2005ki,Cicoli:2007xp,Cicoli:2008va}, although it lately has been criticised in \cite{Junghans:2022exo,Junghans:2022kxg, Gao:2022fdi,Gao:2022uop}.

In type massive IIA string theory, the celebrated four-dimensional DGKT construction \cite{DeWolfe:2005uu} also serves as a counterexample to the strong ADC (however see \cite{Buratti:2020kda} for a modification to the conjecture). There all the moduli are stabilised with classical ingredients such as smeared orientifold-6 (O6-)planes on a $T^6/\mathbb{Z}_3 \times \mathbb{Z}_3$ orbifold. More generally this can be extended to other constructions on manifolds with SU(3)-structure in type IIA. Such solutions were pioneered by \cite{Lust:2004ig}, furthermore studied in references like \cite{Behrndt:2004km,Behrndt:2004mj, Camara:2005dc} and extended in other works, e.g. \cite{Grana:2006kf, Caviezel:2008ik, Lust:2009mb, Lust:2009zb} and many others.  See also \cite{Andriot:2022way,Andriot:2022yyj} for an extensive search of scale-separated solutions amongst type II vacua on group manifolds. The fact that these models have no M-theory uplift and that O6-planes are treated as non-local objects has led to a lot of debate about the consistency of these solutions. Advancements on the latter issue have been made, treating the O6-planes as local sources by using a perturbative scheme \cite{Junghans:2020acz,Marchesano:2020qvg}. Furthermore, a double T-dual version of \cite{DeWolfe:2005uu} has been studied on the Iwasawa nilmanifold in type IIA \cite{Cribiori:2021djm}. These solutions have a separation of the KK and AdS scale, but not of the curvature scales as expected from \cite{Tsimpis:2012tu, Andriot:2018tmb} and hence circumventing the nogo-theorem formulated in \cite{Gautason:2015tig}. The strongly coupled solutions of \cite{Cribiori:2021djm} can then further be uplifted to M-theory. Such M-theory solutions also exhibit scale separation and predict the existence of seven-dimensional positively curved Einstein manifolds where the Kaluza-Klein length scale also decouples from the curvature scale. Thus the uplift circumvents the nogo-theorem of \cite{Gautason:2015tig} as well and goes against the claims of \cite{Collins:2022nux}.

The ultimate goal is to settle the debate by the use of holography. Indeed, the CFT$_3$ dual statement of scale separation is quite simple: it requires that the CFT$_3$ has a parametrically large gap. Whether such CFTs exist has not been settled yet, but pieces of the puzzle have been studied in e.g. \cite{Polchinski:2009ch, Alday:2019qrf}. The first steps towards studying the putative CFT$_3$ duals of the models described above can be found in \cite{Conlon:2018vov,Conlon:2020wmc,Conlon:2021cjk, Apers:2022zjx, Apers:2022tfm}. This is often referred to as the `holographic swampland' program. However, CFT$_2$'s are easier to study and hence it would be interesting to test the same statements for AdS$_3$ vacua.
An example of such scale-separated vacua are compactifications of massive type IIA on manifolds with $G_2$-holonomy, studied in \cite{Farakos:2020phe,Farakos:2020idt}. These can be regarded as the 3d analogue of the 4d DGKT solutions. These setups also contain O6-planes, wrapping 4-cycles in the 7d manifold. For some particular choices of the four-form fluxes, scale separation can be achieved in these models as well. A similar construction has been considered in type IIB on manifolds with $G_2$-structure but scale separation could not be obtained \cite{Emelin:2021gzx}.
In order to obtain all these vacuum solutions, the authors had to go through the process of dimensionally reducing the ten-dimensional type IIA/B action, constructing the scalar potential, deriving a 3d real superpotential from that and then minimising this superpotential (in 4d referred to as `solving the F-term equations') to obtain supersymmetric solutions. This was a quite laborious procedure, and the main point of this paper is to demonstrate that such solutions can be found more elegantly by considering the bispinor equations of \cite{Dibitetto:2018ftj}. These techniques were already applied to $\mathcal{N}=1$ vacua with a specific $G_2$-structure in another orientifold setting in \cite{Dibitetto:2018ftj} and to $\mathcal{N}=2$ $G_2$-vacua in \cite{Passias:2020ubv, Macpherson:2021lbr}. In the $\mathcal{N}=1$ setup that is studied in this paper, these bispinors will also be used to construct the 3d superpotential, valid for both type IIA and type IIB.

In section \ref{sec:BispinorGeneralities} the bispinor technology for 3d backgrounds and the properties of a manifold with $G_2$-structure are reviewed. In section \ref{sec:typeIIA} this is applied to type IIA backgrounds, in section \ref{sec:typeIIB} for type IIB ones. Section \ref{sec:superpotential} discusses the 3d superpotential in the bispinor formalism and its variations are shown to be equivalent with the bispinor equations.

\section{Bispinors in 7d}
\label{sec:BispinorGeneralities}
A standard way to find vacuum solutions in string theory is dimensionally reducing the ten-dimensional theory to a lower dimensional effective field theory (EFT) and then minimising the scalar potential $V$ for the moduli. If one insists on keeping some supersymmetry, it is useful to express the scalar potential in terms of a superpotential. A supersymmetric solution is then found by minimising this superpotential.

However, one can alternatively work directly in ten dimensions and find vacuum solutions by solving the 10d vacuum equations of motion. The solutions one obtains in this way are not necessarily supersymmetric either.
Fortunately it has been known that solving the supersymmetry conditions or Killing spinor equations and imposing the Bianchi identities for the RR fluxes is equivalent to solving the equations of motion and guarantees supersymmetry (see \cite{Tomasiello:2022dwe} and references therein for an extensive treatment). Furthermore, these Killing spinor equations can be recast into form- rather than spinor equations, and these are called the bispinor equations. In this formalism, all the data from the metric is encoded into the bispinors. Many works on this technology have appeared in the context of 4d, i.e. \cite{Grana:2004bg,Grana:2005sn,Grana:2006kf,Tomasiello:2007zq,Koerber:2007xk}. More recently, a similar exercise has been made for three dimensions in \cite{Dibitetto:2018ftj} and was applied to many different backgrounds in e.g. \cite{Passias:2019rga,Passias:2020ubv,Couzens:2022agr}. The basics about these setups are reviewed in the remainder of this section, which relies on \cite{Passias:2019rga}.
The 3d compactifications of interest have the following 10d product metric in string frame
\be
    \dd s_{10}^2 = e^{2A}\dd s_3^2 + \dd s_7^2,
\ee
where the warp factor $e^A$ is a function on the 7d space.
The 10d spinors of type IIA/B are therefore accordingly decomposed into a three-dimensional ($\zeta$) and seven-dimensional ($\eta_{1,2}$) part, being
\be
    \epsilon_1 =\zeta \otimes \eta_1 \otimes \begin{pmatrix} 1 \\ -i \end{pmatrix}, \qquad
    \epsilon_2 =\zeta \otimes \eta_2 \otimes \begin{pmatrix} 1 \\ \mp i \end{pmatrix}.
\ee
The upper (lower) sign corresponds to type IIA (IIB) string theory. In a supersymmetric AdS$_3$ background, the external 3d spinor $\zeta$ has to satisfy 
\be
    \nabla_\nu \zeta = \frac{\mu}{2}\gamma_\nu \zeta,
\ee
where $\mu$ parametrises the AdS scale in string units, i.e. $\mu^2 = L_\text{AdS}^{-2}$ and $\Lambda = -2 \mu^2$. Evidently one should take $\mu = 0$ for Minkowski backgrounds.
The supersymmetry conditions that involve these spinors, can be written in terms of polyform equations. Indeed, the bispinor constructed out of $\eta_1$ and $\eta_2$ can be related through the Clifford map to an even polyform $\Psi_+$ and an odd $\Psi_-$ which are often called the bispinors:
\be
    \eta_1 \otimes \eta_2^\dagger = \Psi_+  = i \Psi_-. 
\ee
Unlike the for 6d Euclidean spinors, $\Psi_+$ and $\Psi_-$ in 7d are related to each other through the Clifford map, i.e.
\be
    \star \lambda \Psi_\pm = \mp \Psi_\mp.
\ee
The $\star$ is the usual Hodge operator and $\lambda$ represents the reversal operator on a $p$-form with $\lambda A_p = (-1)^{\lfloor \frac{p}{2}\rfloor} A_p$.
These bispinors encode all the information of the 7d internal metric and are subject to the normalisation condition
\be
    \left<\Psi_+,\Psi_- \right> = \frac{e^{2A}}{8} \vol_7,
\ee
where $\vol_7$ is the volume form (not with unit volume). Also, $\left<\Xi, \Upsilon \right> = (\Xi \wedge \lambda\Upsilon  )|_\text{top}$ is the Mukai-Chevally pairing for the two polyforms $\Xi$ and $\Upsilon$. In 7d, the $\star \lambda$- and pairing operator have useful properties:
\be
\label{eq:Hodge&PairingProp}
    \left<\Xi, \Upsilon \right> = -\left<\Upsilon, \Xi \right>, \qquad  \left<\star \lambda \Xi, \star \lambda \Upsilon \right> =  \left<\Xi, \Upsilon \right>, \qquad (\star \lambda)^2 =-1.
\ee
As said before, the supersymmetry conditions for type II theories can be reformulated in terms of the bispinors. It is a laborious procedure, but the result is rather elegant.  For orientifold vacua, for which the two spinors $\eta_{1,2}$ have the same norm, they are \cite{Dibitetto:2018ftj}:
\begin{align}
\label{eq:BPS1}
    &\dd_H( e^{A-\phi}\Psi_{\mp}) =0\\
\label{eq:BPS2}
    &\dd_H(e^{2A-\phi}\Psi_{\pm}) \mp 2 \mu e^{A-\phi}\Psi_{\mp} = \frac{1}{8}e^{3A}\star \lambda F\\
\label{eq:BPS3}
    &\left<\Psi_{\mp}, F \right> = \mp \frac{\mu}{2}e^{-\phi} \vol_7,
\end{align}
where the upper (lower) sign has to be taken for type IIA (IIB), $\dd_H = \dd - H\wedge$ is the $H$-twisted exterior derivative and the fluxes $F$ have only legs in the 7 internal dimensions.
The last equation, \eqref{eq:BPS3}, is referred to as the pairing equation and is absent for 4d supergravities.
Supplemented with the RR Bianchi identities, 
\be
    \dd_H F = j_{\text{source}},
\ee
the bispinor equations solve the 10d equations of motion. The beauty of this system of equations is that the local sources like D-branes or O-planes only directly affect the Bianchi identities and not the bispinor equations. However, since O-planes do project out certain fluxes and forms, one has to make sure that the chosen orbifold and orientifold involutions are compatible with the bispinors $\Psi_\pm$ of a given group structure.

In the case of interest, being a 7d manifold with $G_2$-structure, the bispinors are \cite{Dibitetto:2018ftj, Passias:2019rga}
\begin{equation}
\label{eq:Psi+-}
    \Psi_- = \frac{e^A}{8} \left( \Phi - \vol_7\right), \qquad 
    \Psi_+ = \frac{e^A}{8} \left( 1 - \star \Phi\right),
\end{equation}
where $\Phi$ is the $G_2$-invariant three-form. Together with its associate four-form $\Psi = \star \Phi$ it can be used to reconstruct the metric by \cite{Tomasiello:2022dwe}
\be
\label{eq:G2metric}
    g_{mn} = \frac{1}{24} \Phi_{mpq}\Phi_{nrs} \Psi^{prqs}.
\ee
The last object $\Psi^{prqs}$ has to be seen as a `quadruvector' whose elements are defined by $\Psi^{pqrs} = - (\Psi_{pqrs})^{-1}$. Both the three- and four-form can be parametrised in terms of the vielbein one-forms $e^a$, $a=1...7$ as 
\begin{equation}
\begin{alignedat}{1}
    \Phi &= e^{127} + e^{347} + e^{567} + e^{145} + e^{136} + e^{235} - e^{246}\\
    \star \Phi &= e^{3456} + e^{1256} + e^{1234} + e^{2367} + e^{2457} + e^{1467} - e^{1357}.
\end{alignedat}
\end{equation}
Whenever the three-form and its associate four-form are not closed, their departure from non-closure can be parametrised by torsion classes $W_j$ which are classified in in the $G_2$-irreducible representations. These comprise the trivial \textbf{1}-, the fundamental \textbf{7}-, the adjoint \textbf{14}- and symmetric traceless \textbf{27}-representations which correspond to the torsion classes $W_1$, $W_7$, $W_{14}$ and $W_{27}$ respectively. Such torsion classes are often referred to as geometric flux. In summary, non-closure of the three- and four-form is given by
\begin{align}
\label{eq:G2_Torsion}
\begin{split}
    &\dd \Phi = W_1 \star \Phi + W_7 \wedge \Phi + W_{27}\\
    &\dd \star \Phi = \frac{4}{3} W_7 \wedge \star \Phi + W_{14} \wedge \Phi.
\end{split}
\end{align}
The $W_1$- and $W_7$-torsion classes are a 0- and 1-form respectively, and the $W_{14}$ and $W_{27}$ are a 2- and 4-form that satisfy
\be \label{eq:torsion_properties}
    \star W_{14} = -W_{14} \wedge \Phi, \qquad W_{14} \wedge \star \Phi = 0, \qquad W_{27} \wedge \Phi = 0 = W_{27} \wedge \star \Phi.
\ee
We now established the necessary tools to discuss AdS$_3$ solutions with $G_2$-structure, alike \cite{Farakos:2020phe, Emelin:2021gzx}, in the language of the bispinors. This will be discussed in the next sections.

\section{Type IIA \texorpdfstring{AdS$_3$}{AdS3} solutions}
\label{sec:typeIIA}
In this section we look at AdS$_3$ backgrounds in type IIA with bispinor techniques. Some of these setups were previously considered in \cite{Farakos:2020phe}. A beautiful feature of the bispinor language is that the bispinor equations leave us with conditions on the fluxes and geometric forms for which the presence of sources like O-planes is not directly invoked. Besides the requirement that they have to be  compatible with a $G_2$-structure and preserve the right amount of supersymmetry, the sources only play an explicit role in the Bianchi identities. This allows us to make general statements about the solution without specifying the sources at the beginning.

When we consider type IIA supergravity, the first bispinor equation \eqref{eq:BPS1} with the $G_2$-structure expressions for $\Psi_\pm$ \eqref{eq:Psi+-} implies that
\begin{align}
\label{eq:PhiTorsion}
    &\dd \left(e^{2A-\phi}\Phi\right) = 0, \qquad \Phi \wedge H = 0.
\end{align}
On top of that, by the virtue of \eqref{eq:BPS2} the RR fluxes appear to be
\begin{align}
\label{eq:F6BPS}
    &e^{3A}\star F_6 = -\dd \left(e^{3A-\phi}\right) \\
\label{eq:F4BPS}
    &e^{3A} \star F_4 = - e^{3A-\phi} H - 2 \mu e^{2A-\phi} \Phi\\
\label{eq:F2BPS}    
    &e^{3A} \star F_2 = \dd\left(e^{3A-\phi} \star \Phi \right)\\
\label{eq:F0BPS}    
    &e^{3A} \star F_0 = e^{3A-\phi}H \wedge \star \Phi + 2 \mu e^{2A-\phi} \vol_7
\end{align}
The pairing equation \eqref{eq:BPS3} puts additional constraints on the fluxes, i.e.
\begin{align}
\label{eq:PairingBPS}
    \Phi \wedge F_4 - F_0 \vol_7 = -4 \mu e^{-A-\phi} \vol_7.
\end{align}
Furthermore, it is interesting to combine eq. \eqref{eq:F4BPS}, \eqref{eq:F0BPS} and \eqref{eq:PairingBPS} to obtain
\begin{equation}
\label{eq:HandF0simplified}
    H \wedge \star \Phi = -6\mu e^{-A} \vol_7, \qquad e^{\phi}F_0 = -4 \mu e^{-A}.
\end{equation}
From the latter condition we derive that $\e^{\phi+A}$ is a constant, since $F_0$ needs to be a quantised constant. More importantly we conclude that \\
\textit{There are no supersymmetric AdS$_3$ compactifications of massless type IIA string theory with $G_2$-structure.}\\
This is because massless type IIA needs $F_0=0$ and that requires $\mu=0$ through eq. \eqref{eq:HandF0simplified}, leaving us only with Minkowski solutions in massless type IIA.
For the 4d DGKT vacua in massive type IIA on manifolds with SU(3)-structure it was possible to perform two T-dualities and end up with vacua in massless type IIA, again on manifolds with SU(3)-structure \cite{Caviezel:2008ik, Cribiori:2021djm}. This scenario is therefore not possible in 3d vacua from manifolds with $G_2$-structure.

In what follows we are mainly interested in setups with smeared sources.
For those the warping and dilaton profile are trivial. From eq. \eqref{eq:PhiTorsion}, this means that all torsion classes except $W_{14}$ vanish:
\be
    \dd \Phi = 0, \qquad \dd \star \Phi = W_{14}\wedge\Phi.
\ee
These are the so-called closed $G_2$-structures. A manifold with such $W_{14}$ torsion cannot be a toroidal orbifold \cite{DallAgata:2005zlf}, but it can be realised as a compact solvmanifold \cite{Fernandez:2018aaa, Fino:2021}. By \eqref{eq:F2BPS} the inclusion of $W_{14}$ also means that  $F_2 = -e^{-\phi} W_{14}$. 

Smeared O-planes have to be introduced to solve the Bianchi identities, and hence the equations of motion. In its most general form, they are:
\be
\label{eq:IIA_BianchiIdentity}
\begin{alignedat}{2}
    &\dd F_0 = 0, \qquad
    &&\dd F_2 = H F_0 + j_\text{O6/D6},\\
    &\dd F_4 = H \wedge F_2 + j_\text{O4/D4} \qquad
    &&\dd F_6 = H \wedge F_4 + j_\text{O2/D2}.
\end{alignedat}
\ee
In the setups studied in the next subsection, $W_{14}=0$ and there are no calibrated two-cycles that spacetime-filling O4-planes or D4-branes can wrap \cite{Fino:2021}.

From now on we write $e^{\phi}=g_s$ and put $e^A = 1$ without loss of generality for the rest of this section.
One remarks from \eqref{eq:F6BPS} that with these quantities being constant, $F_6$ necessarily vanishes. 
\subsection*{A toroidal orientifold example and scale separation}
As a concrete example we consider a toroidal orientifold $T^7/\mathbb{Z}_2 \times \mathbb{Z}_2\times \mathbb{Z}_2$ with $G_2$-structure as was studied in \cite{Farakos:2020phe}. This is a seven-dimensional orbifold that has orbifold singularities that can be chosen to be blown up. For such toroidal orbifolds it is known that the torsion class $W_{14}$ vanishes as stated before \cite{DallAgata:2005zlf}. The three orbifold actions can be parametrised as follows:
\begin{align}
    &\Theta_\alpha \(e^1, e^2,e^3,e^4,e^5,e^6,e^7\)  = \(-e^1,-e^2,e^3,e^4,-e^5,-e^6,e^7\) \\
    &\Theta_\beta \(e^1,e^2,e^3,e^4,e^5,e^6,e^7\)  = \(e^1,e^2,-e^3,-e^4,-e^5,-e^6,e^7\) \\
    &\Theta_\gamma \(e^1,e^2,e^3,e^4,e^5,e^6,e^7\)  = \(-e^1,e^2,-e^3,e^4,-e^5,e^6,-e^7\).
\end{align}
There are 7 sets of O6-planes that are generated by these orbifold actions and the following orientifold action:
\be
    \sigma\(e^1, e^2,e^3,e^4,e^5,e^6, e^7\)  = \(-e^1, -e^2,-e^3,-e^4,-e^5,-e^6,-e^7\).
\ee
It also generates spacetime-filling O2-planes.
The O6-planes source the $F_2$ Bianchi identity, the O2-planes the one of $F_6$ as in \eqref{eq:IIA_BianchiIdentity}. We allow for D2-branes to potentially cancel the O2-charge. 

For the remainder of this section, it is useful to write the fluxes and geometric forms in terms of the three-form basis $\{\Phi_i\}$ and four-form basis $\{\Psi_i\}$ for $i=1,...,7$ that satisfy
\be
    \Phi_i \wedge \Psi_j = \delta_{ij} \tilde{\star}1, \qquad \Phi_i \wedge \Phi_j = 0, \qquad \Psi_i \wedge \Psi_j =0.
\ee
Unit volume form is understood for $\tilde{\star}1$. The $G_2$-invariant three-form can then be decomposed in the three-form basis as
\be
    \Phi = s^i \Phi_i,
\ee
where the $s_i$ are the geometric moduli which can be reformulated into the radii of the different circles in the $T^7$. With these moduli, another useful identity is 
\be
      \star \Phi_i = \frac{V_7}{(s^i)^2} \Psi_i,
\ee
with $V_7 = \(\Pi_{i=1}^7 s^i\)^{1/3}$ the volume.
The fluxes can similarly be decomposed in the three- and four-form basis:
\be
    H = h^i \Phi_i, \qquad F_4 = F_{4A} + F_{4B} = f_A^i \Psi_i + f_B^i\Psi_i,
\ee
with $F_{4A} \wedge H = 0, F_{4B} \wedge H \neq 0$. In other words, $F_{4A}$ does not participate in the $F_6$ Bianchi identity whereas $F_{4B}$ does.
With all of these redefinitions, the bispinor equations become the algebraic conditions
\begin{align}
\label{eq:F4_Orbifold}
   & g_s (f_A^i + f_B^i) \frac{(s^i)^2}{V_7} = -h^i - 2 \mu s^i \quad \text{no sum}\\
\label{eq:F0_Orbifold}
   & g_s F_0 = \sum_i \left(\frac{h^i}{s^i}\right) + 2\mu \\
\label{eq:pairing_Orbifold}
   & g_s \sum_i(f_A^i + f_B^i) s^i - g_s F_0 V_7= -4\mu V_7.
\end{align}
It is now time to look whether the equations above exhibit a scaling symmetry with the unconstrained four-form flux quanta.
Let us simplify by assuming that all the moduli $s^i$ scale in the same manner, which we will refer to as uniform scalings. Then $s^i \sim V_7^{3/7}$. We make the same assumption for the $h_i$'s, $f_A^i$'s and $f_B^i$'s.
We know that $F_0$ should not scale. Then from the $F_2$ Bianchi identity we know that the $h^i$'s should not scale either, since the O6 source cannot scale. This then means that the $f_B^i$'s can scale neither, seen from the $F_6$ Bianchi identity. By investigation of the equations above, we notice that there is no scaling symmetry unless all the $f_B^i$ vanish, and hence $F_4$ does not participate in the $F_6$ Bianchi identity. But the O2-planes and D2-branes do, so we can only allow for setups with eight times more O2s than D2s such that their total charge vanishes. With this requirement, a scaling symmetry of the bispinor equations and Bianchi identities can be found with a new scaling parameter $f_A$:
\be
\begin{alignedat}{3}
    & f_A^i \sim f_A, \qquad  && F_0 \sim f_A^0, \qquad  && h^i \sim f_A^0.\\
    & g_s \sim f_A^{-3/4}, \qquad && \mu \sim f_A^{-3/4}, \qquad && s^i \sim f_A^{3/4}.
\end{alignedat}
\ee
When we compare the KK-scale to the AdS scale, we find 
\be
    \frac{L_\text{KK}^2}{L_\text{AdS}^2} =V_7^{2/7} \mu^2 \sim f^{-1}
\ee
which goes to zero in the large four-form flux limit and hence scale separation is achieved once a solution can be found.

Let us now look at the same example as in \cite{Farakos:2020phe} with the following flux choices
\begin{align}
    &\tilde{f}^i =f_A(-1,-1,-1,-1,-1,-1,6), \qquad h^i = h(1,1,1,1,1,1,1).
\end{align}
This leads to a six-fold symmetry for the $s^i$ moduli in the first six directions, but not the seventh. This is parametrised by 
\be
    s^{i} = (s,s,s,s,s,s,s_7).
\ee
With the introduction of the following non-scaling quantities
\be
\label{eq:redefsIIAParams}
\begin{alignedat}{2}
    & m = \frac{g_s F_0}{2\mu}, \qquad  &&f = \frac{h g_s f_A }{(2\mu)^2 V_7}.\\
    & x = 2\mu \frac{s}{h}, \qquad  &&y = 2\mu \frac{s_7}{h},
\end{alignedat}
\ee
the algebraic equations \eqref{eq:F4_Orbifold}-\eqref{eq:pairing_Orbifold} become the following simple set of equations
\be
\begin{alignedat}{2}
    & 1+ x- f x^2    = 0, \qquad   && (m-1)x y-6y -x  = 0.\\
    & 1+ y + 6 f y^2  =0, \qquad && 6f(y-x) -m +2 = 0.
\end{alignedat}
\ee
This system can be solved explicitly, with solutions:
\be
\begin{alignedat}{2}
    & m = -2, \qquad && f = \frac{1}{49}\(-10+ \sqrt{2}\)\\
    & x = -3 - \sqrt{2}, \qquad && y = \frac{1}{3} \(1- 2\sqrt{2}\).
\end{alignedat}
\ee
All these values are negative, and if we want positive $s$ and $s_7$ it means that we have to take the signs of $\mu$, $h$, $F_0$ and $f_A$ appropriately. One can show that these numerical values agree with the solution found in \cite{Farakos:2020phe} after a change to Einstein frame for the dilaton. 

There is no guarantee that a solution can be found for any $F_{4A}$- and $H$-flux choice. It would therefore be interesting to chart the landscape of scale-separated AdS$_3$ vacua in future work.

\section{Type IIB \texorpdfstring{AdS$_3$}{AdS3} solutions}
\label{sec:typeIIB}
One can do the same exercise for type IIB to find AdS$_3$ vacua. For that one should take the same bispinor equations \eqref{eq:BPS1}-\eqref{eq:BPS3} but now with the lower signs. In \cite{Emelin:2021gzx} it was observed that it is hard to achieve scale-separated solutions in the type IIB setup. Unsurprisingly we will obtain the same conclusion with the bispinor techniques in a faster fashion.

From the first bispinor equation \eqref{eq:BPS1} one finds that
\be   
\label{eq:starPhiTorsion}
    \dd (e^{2A-\phi}) =0, \qquad  \dd (e^{2A-\phi} \star \Phi) =0, \qquad  H = 0.
\ee
The RR fields can be determined with the second bispinor equation \eqref{eq:BPS2}:
\begin{align}
    &e^{3A}\star F_1 =  0\\
\label{eq:F3BPS}
    &e^{3A} \star F_3 = - \dd (e^{3A-\phi}\Phi) + 2\mu e^{2A-\phi} \star \Phi\\
    &e^{3A} \star F_5 = 0\\
\label{eq:F7BPS}
    &e^{3A} \star F_7 = - e^{2A-\phi} 2\mu.
\end{align}
Here eq. \eqref{eq:starPhiTorsion} was already used to arrive at these expressions. Some of these RR fields are subject to another constraint through the pairing equation \eqref{eq:BPS3}:
\be
\label{eq:IIBPairing}
    \star \Phi \wedge F_3 - F_7 = 4 \mu e^{-A -\phi}\vol_7.
\ee
It serves well to combine \eqref{eq:F3BPS}, \eqref{eq:F7BPS} and \eqref{eq:IIBPairing} to find 
\begin{align}
    &\star \Phi \wedge F_3 = 2 \mu e^{-A -\phi}\vol_7, \qquad \Phi \wedge \( e^{-3A} \dd \(e^{3A - \phi} \Phi\) \) = 12 \mu e^{-A -\phi}\vol_7.
\end{align}
From the latter equation the torsion class $W_1$ can be determined: 
\be
    W_1 = \frac{12}{7} \mu\: e^{-A}.
\ee
Again, we are interested in smeared sources such that the warping and dilaton profile are trivial and we therefore put $e^{\phi}=g_s$ and $e^A=1$. Since this implies $\dd \star \Phi =0$ we find $W_7 = 0 = W_{14}$ due to eq. \eqref{eq:G2_Torsion}. Then we can write $\dd \Phi = W_1 \star \Phi + W_{27}$, which results for \eqref{eq:F3BPS} in
\begin{align}
\label{eq:F3WithTorsion}
    g_s F_3 = \frac{2\mu}{7} \Phi - \star W_{27}. 
\end{align}
The bispinor equations have to be supplemented with the Bianchi identities here as well.
Because $H$, $F_1$ and $F_5$ are identically vanishing but $F_3$ not, there is only the potential need of O5/D5 sources when solving the Bianchi identities and hence only those sources are considered. 
Therefore, the only non-trivial Bianchi identity is the one for $F_3$:
\begin{equation}
    \dd F_3 = j_\text{O5/D5}.
\end{equation}
When looking whether scale separation is possible in these type IIB setups, it is again useful to benefit from a scaling symmetry. Here we only look at setups with uniform scalings, meaning that all volume $p$-forms dual to $p$-cycles scale as $L^p$, where $L = V_7^{1/7}$ is the volume scale and serves as our proxy for the KK scale. This $L$ is then a natural choice to parametrise the scaling symmetry.
By examining the scaling properties of the solutions, some conclusions of \cite{Emelin:2021gzx} are revisited. A first remark is that we need configurations where $F_3$ is closed and thus the D5-branes cancel the O5-charge in order to allow for weakly coupled solutions. Indeed, if not the $F_3$ Bianchi identity tells us that $F_3$ cannot scale. 
We learn further that with uniform scalings $\Phi \sim L^3$, $\star \Phi \sim L^4$ and hence through the torsion equation $\dd \Phi = W_1 \star \Phi + W_{27}$ that $\mu \sim W_1 \sim L^{-1}$ and $W_{27}\sim L^3$ or $\star W_{27} \sim L^2$ when respecting geometric flux quantisation. All these scalings indicate that the RHS of \eqref{eq:F3WithTorsion} scales like $L^2$, and hence $g_s \sim L^2$ because $F_3$ does not scale. Thus a large volume limit forces us to a strongly coupled regime. Therefore we need the D5-charge cancelling the O5-charge such $\dd F_3 =0$ is enforced. Taking the easy route and requiring $F_3$ to vanish would lead us to Minkowski compactifications and hence are not of interest here.
So in the case $F_3 \neq 0$ and $\dd F_3 =0$, we conclude from above that 
\be
    \mu \sim L^{-1}, \qquad    g_s \sim L^{2-a}, \qquad F_3 \sim L^{a},
\ee
with $a>2$.
However, scale separation cannot be achieved, because the ratio
\be
    \frac{L_\text{KK}^2}{L_\text{AdS}^2} = L^2 \mu^2  \sim L^0
\ee
is of order one and cannot be made parametrically small. Hence there are no scale-separated AdS$_3$ solutions of type IIB on manifolds with $G_2$-structures for uniform scalings, just as was observed in \cite{Emelin:2021gzx}. The fact that one cannot find such scale-separated 3d vacua with weak coupling and large volume in classical type IIB is quite interesting. Similarly, in \cite{Cribiori:2021djm} it was observed that, despite earlier claims, the classical type IIB AdS$_4$ vacua with O5/O7-planes of \cite{Petrini:2013ika} and \cite{Caviezel:2009tu} do not admit scale separation, weak coupling and large cycles at the same time. Therefore one could suggest that classical type IIB string theory does not harbour such vacuum solutions.

\section{The 3d superpotential}
\label{sec:superpotential}
The type II solutions discussed above benefit from $\mathcal{N}=1$ supersymmetry in 3d. Just as in the 4d case, the scalar sector of such a supergravity theory is governed by a superpotential. For these type II orientifolds on manifolds with $G_2$-structure, the real superpotential $\mathcal{P}$ was computed for type IIA and type IIB in \cite{Farakos:2020phe} and \cite{Emelin:2021gzx} separately. Alternatively one can define a normal superpotential $\mathcal{W}$ and a K\"ahler-like potential $\mathcal{K}$ (just as in 4d) from this real superpotential through the relation $\mathcal{P} = \left|e^{\mathcal{K}/2} \mathcal{W} \right|$. In 4d, the superpotential is the celebrated Gukov-Vafa-Witten superpotential and extensions thereof \cite{Gukov:1999ya,Grimm:2004ua,Grimm:2004uq}. Moreover, in \cite{Koerber:2007xk} the 4d off-shell K\"ahler- and superpotential were determined using the language of bispinors. This has the benefit that expressions for type IIA and IIB are the same upon the exchange of the odd and even bispinors. In what follows it is argued that the same can be done for these 3d setups with $G_2$-structure.

\noindent To put it simply, the 3d superpotential $\mathcal{W}$ and K\"ahler potential $\mathcal{K}$ are
\begin{align}
\label{eq:superpot}
    \mathcal{W} &= \pm 2\int \left<e^{A-\phi}\Psi_\mp, F \right> - 8 \int e^{-4A}\left<\dd_H \( e^{2A-\phi}\Psi_\pm \), e^{2A-\phi}\Psi_\pm \right>\\
\label{eq:kahlerpot}
    \mathcal{K} &= - 4 \log \left[8\int e^{-A -2\phi}\left<\Psi_+, \Psi_- \right>\right] 
\end{align}
It is important to note that in 3d there is no notion of complex moduli fields, and hence the superpotential we have written down is not to be understood as a holomorphic function anymore. Furthermore, it was shown in the previous sections that the warping should be trivial for the models under consideration. However, the warp factors are kept explicitly to make contact with the bispinor equations \eqref{eq:BPS1}-\eqref{eq:BPS3}. The expression for the K\"ahler potential is similar to its 4d counterpart in \cite{Koerber:2007xk}. With the constant profiles, the K\"ahler potential can be neatly written as
\be
    \mathcal{K} = - 4 \log \[e^{A-2\phi} V_7\] = - 4 \log \[e^{A} V_7\] + 8 \phi.
\ee
The super- and K\"ahler potential given here are in agreement with \cite{Farakos:2020phe,Emelin:2021gzx} after a change to Einstein frame for the dilaton.

One can show that the usual supersymmetry conditions for this K\"ahler- and superpotential are equivalent to the bispinor equations \eqref{eq:BPS1}-\eqref{eq:BPS3}. With the real moduli $\varphi_i$, these conditions imply
\be
    0= D_{\varphi_i} \mathcal{W} \equiv \frac{\delta}{\delta\varphi_i} \(e^{\frac{\mathcal{K}}{2}}\mathcal{W}\) = \frac{\delta \mathcal{W}}{\delta\varphi_i}  + \frac{1}{2}\frac{\delta\mathcal{K}}{\delta\varphi_i}\mathcal{W}_\text{on-shell}.
\ee
The on-shell superpotential $\mathcal{W}_\text{on-shell}$ has to satisfy $\Lambda = - 8 e^{\mathcal{K}}|\mathcal{W}_\text{on-shell}|^2$, which then can be derived from eq. \eqref{eq:kahlerpot} to be
\be
\label{eq:on-shell_superpot}
    \mathcal{W}_\text{on-shell} = \frac{\mu}{2} e^{A - 2\phi} V_7.
\ee
Alternatively, one gets the same result when plugging in the bispinor equations \eqref{eq:BPS1}-\eqref{eq:BPS3} into \eqref{eq:superpot}.

With the expression for the on-shell superpotential at hand, one can show that the supersymmetry condition for the dilaton $D_\phi \mathcal{W} =0$ is equivalent to the pairing equation \eqref{eq:BPS3}:
\begin{align}
\nonumber 
    0 &= D_\phi W \\
   \Rightarrow 0 &=
    \left<e^{A-\phi}\Psi_\mp, F \right> \mp 8e^{-4A}\left<\dd_H \(e^{2A-\phi}\Psi_\pm\) ,  e^{2A-\phi}\Psi_\pm \right> \mp2 \mathcal{W}_\text{on-shell}\frac{\vol_7}{V_7}\\
   &= -\left<e^{A-\phi}\Psi_\mp, F \right> \mp  \mathcal{W}_\text{on-shell}\frac{\vol_7}{V_7},
\end{align}
and hence with \eqref{eq:on-shell_superpot} we recover the pairing equation \eqref{eq:BPS3}
\be
\label{eq:PairingFromSuperpot}
     \left<\Psi_\mp, F \right> = \mp \frac{\mu}{2}e^{-\phi}\vol_7.
\ee

The first bispinor equation \eqref{eq:BPS1} is obtained by requiring that the K\"ahler- and superpotential are invariant under deformations of the RR gauge potential $C$. Consider such an infinitesimal deformation acting as $F\to F + \dd_H \delta C$ with arbitrary gauge potential $\delta C $. Invariance of the superpotential requires
\be
    0 = \delta \mathcal{W} = \pm 2 \int \left< e^{A-\phi}\Psi_\mp , \dd_H \delta C \right> = \mp 2 \int \left< \dd_H \(e^{A-\phi}\Psi_\mp\) , \delta C \right>. 
\ee
Integration by parts was used in the last step.
Because $\delta C$ is arbitrary, we recover the first bispinor equation \eqref{eq:BPS1}:
\be
\label{eq:derivedBPS1}
    \dd_H \( e^{A-\phi} \Psi_\mp \) = 0.
\ee
It can also be obtained by acting with $\dd_H$ on the second bispinor \eqref{eq:BPS2} and requiring the equations of motion for the RR fluxes $F$.

The second bispinor equation \eqref{eq:BPS2} can be derived from deformations of the bispinors in the super- and K\"ahler potential. For that it is useful to classify all possible forms into the irreducible representations of the $G_2$-group. These consist of the trivial \textbf{1}-, the fundamental \textbf{7}-, the adjoint $\textbf{14}$- and the symmetric traceless \textbf{27}-representation. The spaces of $p$-forms $\Lambda^p$ are categorised under these representations as follows
\begin{equation}
    \begin{alignedat}{1}
         \Lambda^0, \: \Lambda^7 &: \quad {\bf 1}\\
        \Lambda^1, \: \Lambda^6 &:  
        \quad {\bf 7}\\
        \Lambda^2, \: \Lambda^5 &: \quad {\bf 7} \oplus {\bf 14}\\
         \Lambda^3, \: \Lambda^4 &: \quad {\bf 1} \oplus{\bf 7} \oplus {\bf 27}.
    \end{alignedat}
\end{equation}
The bispinors $\Psi_\pm$ both belong to the trivial representation {\bf 1}. One can then define projectors $\pi^{\bf i}$ that project a form to its component along the representation {\bf i} = {\bf 1, 7, 14, 27}. With such a projector any two polyforms $\Xi$ and $\Upsilon$ satisfy
\be
\label{eq:reprMukaiPairing}
    \left<\pi^{\bf i}(\Xi), \pi^{\bf j}(\Upsilon)\right>= 0 \quad \text{for } {\bf i} \neq {\bf j}.
\ee
The deformations of the bispinors themselves can thus also be categorised in these representations. Since the bispinors encode all the metric information, deforming them also means deforming the metric and hence the Hodge dual operator. We then write such a deformation as $\Psi_\pm \to \Psi_\pm' = \Psi_\pm + \delta_{\bf i}\Psi_\pm$ and $\star \to \star' = \star + \star^{(\delta)}$ where $\star$ still corresponds to the undeformed Hodge star. The bispinors are still dependent on each other and hence we require that $\star' \lambda \Psi_\pm' = \mp \Psi_\mp'$. To first order in the deformation this means that 
\be
    \star^{(\delta)} \lambda \Psi_\pm + \star \lambda \(\delta_{\bf i}\Psi_\pm\) = \mp \delta_{\bf i} \Psi_\mp.
\ee
It is worth to notice that generally the first term, corresponding to deformations of the Hodge operator, does not need to vanish. For the {\bf 1}- and {\bf 27}-deformations this turns out to be the case. For {\bf 7}-deformations, the first term needs to vanish, since such deformations involve 1-, 2-, 5-, and 6-forms, which the deformed Hodge operator cannot generate when acting on the bispinors since these contain only 0-, 3-, 4- and 7-forms. Moreover, there are no {\bf 14}-deformations of the bispinors.

\subsubsection*{1- and 27-deformations}

A deformation in the trivial representation {\bf 1} corresponds to a uniform volume rescaling of all the $G_2$-structure forms. An infinitesimal deformation parametrised by $\delta L$ acts on the forms as follows:
\be
    \delta_{\bf 1} \Phi = 3\: \delta L\:\Phi, \qquad \delta_{\bf 1} \(\star \Phi\) = 4 \:\delta L\:\star \Phi, \qquad  \delta_{\bf 1}\vol_7 = 7\: \delta L \: \vol_7.
\ee
With these we can verify that
\begin{equation}
\label{eq:propTrivialDeform}
    \delta_{\bf 1} \Psi_\mp = \pm \star \lambda (\delta_{\bf 1} \Psi_\pm) + 7\: \delta L\: \Psi_\mp, \qquad 7 \: \delta L\:\vol_7 = \mp 16 e^{-2A} \left<\Psi_\mp, \delta_{\bf 1} \Psi_\pm \right>.
\end{equation}
The supersymmetry condition $\delta_{\bf 1} \mathcal{W} + (\mathcal{W}/2)\delta_{\bf 1}K =0$ for such a deformation then results in 
\begin{equation}
\label{eq:RAW1deform}
    0=
    \mp \left<e^{A-\phi}\delta_{\bf 1}\Psi_\mp, F \right> + 8e^{-4A}\left<\dd_H \(e^{2A-\phi}\Psi_\pm\) ,  e^{2A-\phi}\delta_{\bf 1}\Psi_\pm \right> +  \frac{\mathcal{W}_\text{on-shell}}{V_7}(7 \: \delta L\:\vol_7).
\end{equation}
The first term can now be rewritten with eq.  \eqref{eq:PairingFromSuperpot} and \eqref{eq:propTrivialDeform}, and one finds after multiplying with $e^{2A+\phi}$ that
\begin{align}
\nonumber
    0=
    &-\left<\star \lambda (\delta_{\bf 1}\Psi_\pm), e^{3A}F \right> + 8\left<\dd_H \(e^{2A-\phi}\Psi_\pm\) , \delta_{\bf 1}\Psi_\pm \right>\\ &+\(\frac{\mu}{2}e^{3A-\phi}
    + e^{2A+\phi} \frac{\mathcal{W}_\text{on-shell}}{V_7}\)(7 \: \delta L\:\vol_7).
\end{align}
Now using \eqref{eq:Hodge&PairingProp}, \eqref{eq:on-shell_superpot} and \eqref{eq:propTrivialDeform}, we arrive at
\begin{equation}
    0=
    \left<-\frac{1}{8}e^{3A}\star \lambda F + \dd_H\( e^{2A-\phi}\Psi_\pm\)  \mp 2\mu e^{A-\phi}\Psi_\mp, \delta_{\bf 1}\Psi_\pm \right>.
\end{equation}
In order for this to vanish, the \textbf{1}-component of the first argument of the pairing needs to vanish because of eq. \eqref{eq:reprMukaiPairing}. Similarly, one could manipulate \eqref{eq:RAW1deform} such that $\delta_{\bf 1} \Psi_\mp$ appears in the second argument of the pairing. Together, the two conditions reproduce the {\bf 1}-component of the bispinor equation \eqref{eq:BPS2}.

Let us now look at deformations in the {\bf 27}-representation. Such deformations can be generated by a traceless symmetric matrix $\beta_{m} {}^{n}$ through
\be
    \delta_{\bf 27} \Psi_\pm = \beta_{m} {}^{n} e^m \wedge \iota_{n} \Psi_\pm,
\ee
with $\iota_n = e^n \lrcorner$ is a contraction operator. 
Such deformations satisfy
\be
\label{eq:propSymmDeform}
    \star \lambda \(\delta_{\bf 27} \Psi_\pm\) = \pm \delta_{\bf 27} \Psi_\mp.
\ee
The volume is only subject to deformations in the trivial representation, and hence the supersymmetry condition becomes $\delta_{\bf 27}\mathcal{W} = 0$. This is equivalent to 
\be
    0 = \mp \left<e^{A-\phi}\delta_{\bf 27}\Psi_\mp, F\right> + 8 e^{-4A} \left<\dd_H \(e^{2A-\phi}\Psi_\pm\), e^{2A-\phi}\delta_{\bf 27} \Psi_\pm \right>.
\ee
By using \eqref{eq:Hodge&PairingProp} and \eqref{eq:propSymmDeform} this condition is simplified to:
\be
\label{eq:27BPS2}
    0 = \left<-\frac{1}{8}e^{3A} \star \lambda F + \dd_H \(e^{2A-\phi}\Psi_\pm\) , \delta_{\bf 27} \Psi_\pm \right>.
\ee
Again, this has to vanish for all deformations $\delta_{\bf 27}\Psi_\pm$, such that we recover the {\bf 27}-component of the bispinor equation \eqref{eq:BPS2}.

\subsubsection*{7-deformations}
One could wonder about the deformations of the superpotential in the fundamental {\bf 7}-representation. Such deformations do not affect the volume and hence the K\"ahler potential neither. As argued above, those deformations also do not affect the Hodge operator and thus satisfy 
\begin{equation}
    \star \lambda \(\delta_{\bf 7}\Psi_\pm\) = \mp \delta_{\bf 7} \Psi_\mp.
\end{equation}
The effect of such deformations on the superpotential is similar to the {\bf 27}-case, but with a sign difference:
\be
\label{eq:714BPS}
    0 = \left<+\frac{1}{8}e^{3A} \star \lambda F + \dd_H \(e^{2A-\phi}\Psi_\pm\) , \delta_{\bf 7} \Psi_\pm, \right>\,.
\ee
This comes with the wrong sign when compared to \eqref{eq:BPS2}. Does this then invalidate the results? Fortunately it does not by the virtue of the first bispinor equation \eqref{eq:BPS1}, which we independently derived from the superpotential in \eqref{eq:derivedBPS1}. We argued in section \ref{sec:typeIIB} that $H$ and the torsion classes $W_7$ and $W_{14}$ vanish for constant warping and dilaton in type IIB due to the first bispinor equation \eqref{eq:derivedBPS1}. In type IIA we concluded in section \ref{sec:typeIIA} from the same bispinor equation \eqref{eq:derivedBPS1} that $H$ has no {\bf 7}-component (otherwise $\Phi \wedge H \neq 0$) and that $W_7$ vanishes. From all this, we conclude that $\dd_H\( e^{2A-\phi}\Psi_\pm\)$ has no {\bf 7}-component in both type IIA and IIB due to the first bispinor equation we derived in \eqref{eq:derivedBPS1}. Hence, the second term in \eqref{eq:714BPS} vanishes due to eq. \eqref{eq:reprMukaiPairing}. This leaves us with
\be
 0 = \left<\star \lambda F , \delta_{\bf 7} \Psi_\pm \right>\,.
 \ee
 It leads us to conclude that the fluxes cannot have components in the {\bf 7}-representation, i.e.
 \be
    \pi^{\bf 7}(F) = 0\,.
 \ee
This is perfectly consistent with what we have found earlier.

Lastly we remark that a {\bf 7}-deformation is parametrised by an infinitesimal one-form which we call $\delta\alpha_1$. It acts on the bispinors by
\be
    \delta_{\bf 7} \Psi_\pm = \( \delta\alpha_1 \lrcorner - 
    \delta \alpha_1 \wedge \) \Psi_\mp.
\ee
with $\lrcorner$ the contraction operator.
It is interesting to note that such a {\bf 7}-transformation transforms the bispinors of a $G_2$-structure into those of an SU(3)-structure (see e.g. \cite{Witt:2004vr,Dibitetto:2018ftj}). Indeed, for a one-form $v$, one can regard
\be
    \Psi_\pm^\text{SU(3)} = (v\lrcorner - v \wedge ) \Psi_\mp^{G_2}.
\ee
Furthermore, it is a special O(7,7)-transformation that corresponds to T-duality along the vector dual to $v$.

\subsubsection*{\texorpdfstring{$B$}{B}-deformations}
The {\bf 14}-component of the bispinor equations arises by computing variations of the superpotential with respect to the $B$-field. This can be done by writing $F= \dd_H C$ and by considering a variation of $H \to H + \dd (\delta B)$. Aside from the identities~\eqref{eq:Hodge&PairingProp}, the Mukai pairing also satisfies (in 7d):
\begin{equation}\label{eq:Extra_PairingProp}
    \int \left<\dd_H \Xi, \Upsilon\right> = - \int \left<\Xi, \dd_H \Upsilon\right>\,, \quad \left<A\wedge B, C\right> = \left<A, C\wedge \lambda B\right>\,, \quad \lambda(A \wedge B) = \lambda B \wedge \lambda A\,.
\end{equation}
The function $\mathcal{K}$ does not vary under $B$. However, $\mathcal{W}$ does, and both terms in the superpotential take the form
\begin{equation}
    \left<\dd_H \Xi , \Upsilon \right>\,.
\end{equation}
A variation of the B-field generates the following variation:
\begin{align}
    \left<-\dd(\delta B) \wedge \Xi, \Upsilon\right> &= \left<-\dd_H\left(\delta B \wedge \Xi\right)+\delta B \wedge \dd_H\Xi, \Upsilon\right>\\
    &=\left<\delta B \wedge \Xi, \dd_H \Upsilon\right>+\left<\delta B \wedge \dd_H\Xi, \Upsilon\right>\\
    &= \left<\delta B , \dd_H \Upsilon\wedge \lambda\Xi\right>+\left<\delta B , \Upsilon\wedge \lambda\dd_H\Xi\right>\,, \label{eq:B-variation_general}
\end{align}
where eq.~\eqref{eq:Extra_PairingProp} was used.
When $\Xi = \Upsilon = e^{2A-\phi}\Psi_\pm$, this reduces to 
\begin{equation}
    \left<-\dd(\delta B) \wedge e^{2A-\phi}\Psi_\pm, \e^{2A-\phi}\Psi_\pm\right> = 2\left<\delta B , \dd_H\left(e^{2A-\phi}\Psi_\pm\right)\wedge e^{2A-\phi}\lambda\Psi_\pm\right>\,.
\end{equation}
Here, \eqref{eq:Extra_PairingProp} was used again, together with the fact that that only the five-form component in the second argument of the Mukai pairing contributes and hence $\lambda(Z_5) = Z_5$ for a five-form $Z_5$. The same observations can be used for when $\Xi = C$ and $\Upsilon = e^{A-\phi}\Psi_\mp$. The first term in \eqref{eq:B-variation_general} vanishes due to the first bispinor equation, leaving
\begin{equation}
    \left<-\dd(\delta B) \wedge C, e^{A-\phi}\Psi_\mp\right> = \left<\delta B , e^{A-\phi}\Psi_\mp \wedge \lambda F\right>=\left<\delta B , F \wedge e^{A-\phi}\lambda \Psi_\mp \right>
\end{equation}
Combining everything and using $\lambda \Psi_\pm = \pm \Psi_\pm$, we find that
\begin{equation}
    0 = \left<\delta B,  \mp F \wedge e^{A-\phi} \Psi_\mp + 8e^{-4A} \dd_H \left(e^{2A-\phi} \Psi_\pm\right) \wedge e^{2A-\phi} \Psi_\pm\right>
\end{equation}
Using \eqref{eq:reprMukaiPairing}, we find that the {\bf 7}- and {\bf 14}- component of the second argument have to vanish independently. 
For type IIA with constant warping and dilaton ($\exp (A) =1$ and $\exp(\phi) = g_s$), the only components contributing are 
\begin{equation}
    0 = F_2 \wedge \Phi + e^{-\phi}\dd (\star \Phi)\,.
\end{equation}
and so the {\bf 14}-component reads $\pi^\mathbf{14}(F_2) = -e^{-\phi}W_{14}$.
It was already argued above that the {\bf 7}-component, must vanish because the first bispinor equation already tells that $W_{7}$ must vanish. Form this, and \eqref{eq:torsion_properties}, one finds
\begin{equation}
    F_2 = -e^{-\phi}W_{14}\,.
\end{equation}
In type IIB, the relevant components are
\begin{equation}
    0 = -F_1 \wedge \star \Phi + F_5\,.
\end{equation}
It was shown above that the {\bf 7}-components must vanish, so from this one concludes that $\pi^\mathbf{14}(F_5)=0$, and so also $F_5=0$.

\section{Summary}
In this paper it was demonstrated that AdS$_3$ vacua of type II orientifolds with $G_2$-structure can easily be captured by the bispinor techniques of \cite{Dibitetto:2018ftj}. Finding such vacuum solutions with minimal supersymmetry amounts to solving the bispinor equations and the Bianchi identities for the Ramond-Ramond fluxes. This approach leads to the same conclusions as in \cite{Farakos:2020phe,Emelin:2021gzx} for specific types of vacua. The type IIA vacua can be seen as the 3d analogues of the 4d DGKT solutions, they can exhibit scale separation as well and are therefore in tension with the strong ADC \cite{Lust:2019zwm}. It was also argued in \cite{Apers:2022zjx} that they also violate the $\mathbb{Z}_k$-refined strong ADC of \cite{Buratti:2020kda}. 
On the other hand, the type IIB solutions do not exhibit scale separation.

Finally a general formula was given for the 3d superpotential in terms of the $G_2$-structure bispinors and its supersymmetry conditions were shown to be equivalent to the bispinor equations. Although the latter was shown in the context of $G_2$-structures, it is to be expected that the K\"ahler- and superpotential should have very similar expressions for 3d $\mathcal{N}=1$ backgrounds with SU(3)-structure or generalised $G_2$-structure \cite{Witt:2004vr}. T-duality gives a hint for this, as it transforms a $G_2$-structure into a 7d SU(3)-structure but it does not change the final form of the K\"ahler- and superpotential. It would therefore be interesting to derive the 3d superpotential from the gravitino mass, as was done for heterotic 3d supergravity in \cite{deIaOssa:2019cci}. Furthermore, it was shown in \cite{Emelin:2022cac} that the local backreaction of the O6-planes can be computed using the techniques of \cite{Junghans:2020acz} that were developed for 4d DGKT vacua. It would be interesting to do the same exercise with the bispinors of a generalised $G_2$-structure, along the same lines of \cite{Marchesano:2020qvg} where this was done with generalised SU(3)-structures for 4d DGKT solutions. 

\section*{Acknowledgements}
I would like to thank Niall Macpherson, George Tringas, Thomas Van Riet and especially Fotis Farakos and Timm Wrase for enlightening discussions and very useful comments.
This work is supported by grant nr. 1185120N of the Research Foundation - Flanders (FWO) and partially by the KU Leuven C1 grant ZKD1118 C16/16/005.

\bibliographystyle{JHEP.bst}
\bibliography{AllRefs.bib}

\end{document}